\documentclass[a4paper,nofootinbib,twocolumn,showpacs,floatfix,prc]{revtex4}
\usepackage{graphicx}
\usepackage{amsfonts}
\renewcommand{\vec}[1]{\mathbf{#1}}
\def\be{\begin{equation}}
\def\ee{\end{equation}}
\def\bea{\begin{eqnarray}}
\def\eea{\end{eqnarray}}

\begin{document}

\title{Entropy production at freeze-out from dissipative fluids}

\author{E. Moln\'ar$^{1}$}

\affiliation{
$1$ Frankfurt Institute for Advanced Studies, J. W. Goethe University, Max-von-Laue-Str. 1,
D-60438 Frankfurt am Main, Germany
}

\begin{abstract}
{
Abstract: Entropy production due to shear viscosity during the continuous freeze-out
of a longitudinally expanding dissipative fluid is addressed.
Assuming the validity of the fluid dynamical description during the continuous removal of
interacting matter we estimated a small entropy production as function of the freeze-out duration
and the ratio of dissipative to non-dissipative quantities in case of a relativistic massless pion fluid.
}
\end{abstract}


\pacs{24.10.Nz, 25.75.-q}

\maketitle

\section{Introduction}

In the fluid dynamical modeling of high energy heavy-ion reactions the initial large gradients dilute
and cool the matter until the mean free path of particles becomes comparable to the size of the interacting system.
By then the particles do not interact strongly enough to maintain local equilibrium and the
interactions between particles cease leaving the momentum of particles frozen-out \cite{Landau_1}.
\\ \indent
To properly describe the final break-up we need to model the continuous transition from
an interacting fluid to a noninteracting fluid (gas) of particles.
Here we will use the standard first order theory of relativistic
fluid dynamics, i.e., the relativistic Navier-Stokes equations,
and calculate the entropy production during the time the matter
is continuously removed from the interacting fluid.
In reality the matter equilibrates in finite time, where the local relaxation time, $\tau_{th}$,
increases  in time as the system dilutes and the interacting matter turns gradually into a weakly
interacting dissipative fluid before the momentum of particles freeze-out \cite{HiranoGy, Dumitru_0}.
Assuming that the fluid is near local equilibrium during freeze-out we can
estimate the additionally produced entropy from the equations of dissipative fluid dynamics.
Strictly speaking this transition does not necessarily lead to kinetic freeze-out,
since the mean free path of the matter is still finite and the particles may interact further.
The drain from the fluid should be followed by a kinetic description where this transition
can be treated beyond small deviations from local equilibrium, but we will not discuss
the kinetic modeling here.

\section{Freeze-out from dissipative fluids}  \label{DissHydro}

A single-component fluid is characterized by a conserved charge four-current, $N^{\mu}$, the energy-momentum tensor,
$T^{\mu \nu}$, and the entropy four-current, $S^{\mu}$.
The dissipative currents are expressed explicitly by projecting the above quantities perpendicular to the flow of matter \cite{Eckart, Landau_book}.
Hence the conserved charge four-current is decomposed using the transverse projector,
$\Delta^{\mu \nu}= g^{\mu \nu} - u^\mu u^\nu$, where $u^{\mu}$ is the
fluid dynamical flow velocity and $g^{\mu \nu} = \textrm{diag} (1,-1,-1,-1)$ the metric of flat space-time, thus
\be
N^{\mu} = n u^{\mu} + V^{\mu} \, ,
\ee
where $n$ is the local rest frame invariant charge or conserved particle density, and
$V^{\mu} = \Delta^{\mu \nu} N_{\nu}$ is the flow of charge perpendicular to $u^{\mu}$.
\\ \indent
In case of a relativistic massless pion fluid we can neglect the heat conductivity and the bulk viscosity
below the critical temperature since both quantities are very small or vanishing compared to the shear
viscosity \cite{Gavin, Weinberg_book}.
Thus, the energy-momentum tensor is decomposed as:
\be\label{EMtensor1}
T^{\mu \nu} = e u^{\mu} u^{\nu} - p \Delta^{\mu \nu} + \pi^{\mu \nu} \, ,
\ee
where $e$ and $p$ are the local rest frame energy density and isotropic pressure and
the symmetric and traceless part of the energy-momentum tensor is the stress tensor,
$\pi^{\mu \nu} = \left[\frac{1}{2} \left( \Delta^{\mu}_{\alpha} \Delta^{\nu}_{\beta} +
\Delta^{\nu}_{\alpha} \Delta^{\mu}_{\beta} \right) - \frac{1}{3}
\Delta^{\mu \nu} \Delta_{\alpha \beta} \right] T^{\alpha \beta}$.
\\ \indent
These conserved quantities satisfy the following continuity equations written in differential form,
\bea \label{c_conservation}
\partial_{\mu} N^{\mu} &=& 0  \, , \\
\partial_{\mu} T^{\mu \nu} &=& 0 \, ,
\label{em_conservation}
\eea
where, $\partial_{\mu} = \partial/\partial x^{\mu}$, is the space-time four-divergence.
These equations have to satisfy, in addition, the condition of positive entropy production required by
the second law of thermodynamics, thus the entropy production with shear viscosity becomes
\cite{Miklos, Dirk, Muronga, Teaney, Heinz, Romatschke}:
\be\label{entropy_prod_1}
\partial_{\mu} S^{\mu} = \frac{\pi^{\mu \nu} \pi_{\mu \nu}}{2\eta T}  \geq 0 \, .
\ee
Here the shear stress tensor is,
$\pi^{\mu\nu} = 2\eta \sigma^{\mu \nu}$, where $\eta$ is the shear viscosity coefficient and the stress tensor is,
$\sigma^{\mu\nu} = \frac{1}{2} \left( \nabla^\mu u^\nu + \nabla^\nu u^\mu\right)
- \frac{1}{3} \Delta^{\mu\nu} \nabla_\lambda u^\lambda$, and
$\nabla^\mu=\Delta^{\mu\nu} \partial_\nu$.
In the perfect fluid limit, ($\tau_{th} \rightarrow 0$, $\eta \rightarrow 0$), the entropy production is zero \cite{Bebie,Csernai_book}:
\be
\partial_{\mu} S^{\mu}_{perf} \equiv -\frac{\mu}{T} \partial_{\mu} (nu^{\mu}) = 0\, ,
\ee
where $T$ is the temperature, and $\mu$ is the chemical potential.
The above equation is satisfied if the number of charge(s) is conserved or for a vanishing chemical potential.
Aiming to study the freeze-out of massless pions the later condition is favored since $\mu = 0$ \cite{Hung}.
\\ \indent
To describe the freeze-out of matter, one has to separate the fluid into two components \cite{Grassi}:
interacting and frozen-out, (denoted by lower indexes $i$ and $f$).
The interacting component continuously emits particles and its evolution is described via the
equations of fluid dynamics.
The emitted particles leave the fluid and part of it may interact further described by Boltzmann
transport equation \cite{Sinyukov, Bass}, until the particles will no longer collide but
free stream from their kinetic freeze-out points.
\\ \indent
If the particle number is conserved during freeze-out, $\partial_{\mu} N^{\mu} = 0$,
we can separate the system into interacting and frozen-out particles, $N^{\mu} = N^{\mu}_i + N^{\mu}_f$, such that,
\be\label{source_particles}
\partial_{\mu} N^{\mu}_i = - \partial_{\mu} N^{\mu}_f \, .
\ee
Similarly, but independently (even for non-conserved charges), the total energy and momentum
is conserved, $\partial_{\mu} T^{\mu \nu} = 0$, thus the change in the energy-momentum
tensor at freeze-out leads to:
\be \label{source_em}
\partial_{\mu} T^{\mu \nu}_i = -J^{\nu} \, , \qquad \partial_{\mu} T^{\mu \nu}_f = +J^{\nu} \, ,
\ee
where, $J^{\nu} = (J^{0}, \vec{J})$, is the drain term which takes away energy and momentum from
the interacting component and transfers it to the frozen-out component.
Generally there is a non-vanishing momentum transfer, i.e., there is non-zero mechanical work done
on the interacting fluid at freeze-out.
The equation for the entropy, constrained by the second law of thermodynamics is:
\be\label{entropy-current}
\partial_{\mu} S^{\mu} \geq \partial_{\mu} S_i^{\mu} \, ,
\ee
where the entropy production in the frozen-out matter is assumed to be zero.

\subsubsection{Particle number conservation, energy conservation and entropy production}

In what follows, we will work in the light-cone coordinates, making the coordinate transformation
from $(t,z)$ to $(\tau,\eta)$, where $\tau = \sqrt{t^2-z^2}$ is the proper time and
$\eta = \frac{1}{2} \log ((t+z)/(t-z))$ is the space-time rapidity.
Here we present the equations of dissipative fluid dynamics with a drain term using the 1+1D boost
invariant Bjorken model \cite{Bjorken}, where $u^{\mu} = (\cosh\eta, 0,0,\sinh \eta)$ is the flow velocity
and $\theta \equiv \partial_{\mu} u^{\mu} = 1/\tau$,
is the expansion rate independent on the equation of state.
\\ \indent
Assuming that the particle number is conserved leads to the following equation for the interacting density:
\be\label{N_i_1}
dn_i (\tau) = \left[ - d N_i (\tau) - n_i (\tau)\, dV_i(\tau) \right]/V_i(\tau) \, ,
\ee
where the volume of the interacting matter $V_i(\tau) \sim \tau A_T $ increases such as,
$dV_i(\tau)/V_i(\tau) = d\tau/\tau$, and the transverse area of the system is $A_T$.
The volume of the frozen-out particles is unknown from fluid dynamics but it is of no relevance whatsoever,
hence in the following we will not discuss the frozen-out component.
\\ \indent
Furthermore, for simplicity we assume that the change in the number of interacting particles is due
to particle freeze-out only, governed by a boost invariant proper time dependent probability,
$P_{esc}(\tau) d\tau$, such that:
\be\label{dN_i}
dN_i(\tau) = -P_{esc} (\tau) N_i(\tau) d\tau \, ,
\ee
where $dN_f = -dN_i$ and $P_{esc}(\tau)$ is the escape rate.
The properties of matter change considerably in a small space-time domain,
characterized by a time-scale, $\tau'(\tau)$, which governs the process.
Here, we recall the assumption for the escape rate \cite{ModifiedBTE, article_1}, which
characterizes the freeze-out of matter in finite time, $L$, the proper time interval between
two parallel hypersurfaces, where the continuous freeze-out of matter starts at $\tau_0$ and
ends at $\tau_0 + L$, thus
\be\label{escape}
P_{esc}(\tau) \equiv \frac{1}{\tau'(\tau)} = \frac{1}{\tau'(\tau_0)} \frac{L}{L + \tau_0 -\tau} \, ,
\ee
where the parameter, $\tau'(\tau_0)$, denotes the initial characteristic freeze-out time, see Sect. \ref{Results}.
For example at the midrapidity section of the hypersurface, the probability of escape depends
on the invariant distance $x^{\mu} d^3\sigma_{\mu} \sim \tau$ in the direction of
the freeze-out normal vector and on the local properties of matter, such as the proper density and
invariant cross section, i.e., the collision rate, $\Gamma \simeq n \sigma$.
Therefore, $\tau'(\tau) \simeq \Gamma$, is inversely proportional to the relaxation
time \cite{Sinyukov,ModifiedBTE, article_1}.
Fluid dynamics requires that $\Gamma \gg \theta$, while the kinetic freeze-out condition is
$\Gamma \sim \theta$ \cite{Bondorf, FO_Bjorken, Tomasik}.
Now, using the above equations, the change in the interacting particle number density due to simultaneous
Bjorken expansion and freeze-out becomes:
\be\label{dn_i}
\frac{dn_i (\tau)}{d\tau} = - P_{esc}(\tau) n_i(\tau)  - \frac{n_i (\tau)}{\tau} \, .
\ee
Let us recall the first law of thermodynamics neglecting the baryon number and strangeness conservation,
that is, $dE_i = T_i dS_i - p_i dV_i$, where  $p_i dV_i$ denotes the work done by the expansion.
Here, $T_i dS_i = \Delta Q_i + T_i d S^{diss}_i$ denotes the change in the total entropy at constant temperature
and $T_i dS_i \geq \Delta Q_i$ expresses the second law of thermodynamics, where the entropy increase,
$T_i dS^{diss}_i$, is due to dissipation (shear viscosity).
Thus, for an adiabatic expansion, $\Delta Q_i = 0$, the change in the total energy is:
\be\label{pressure_identity}
dE_i(\tau) = -\left[ p_i(\tau) - \Phi_i(\tau)\right] dV(\tau)
\ee
where $p_i(\tau)$ is the isotropic pressure and $\Phi_i(\tau)$ is the stress which slows down the expansion
and decreases the "useful" work done by the pressure.
Furthermore, the pressure is independent on the rapidity, $d p_i(\tau)/d\eta = 0$.
\\ \indent
Now, let us consider simultaneously the adiabatic expansion and the freeze-out of the system.
The decrease in the total energy of the system is due the energy taken away
by the escaping interacting particles and the work done on interacting component by the frozen-out component
during the particle escape from the freeze-out surface \cite{Dumitru_1}.
With a similar ansatz as in eq. (\ref{dN_i}) we obtain an equation for the evolution of the
energy density:
\bea\label{de_i}
\frac{d e_i (\tau)}{d\tau}
&=& - \left[e_i(\tau) + p_i(\tau) \right]P_{esc} (\tau) \\ \nonumber
&-& \frac{ e_i(\tau) + p_i(\tau)}{\tau}  + \frac{\Phi_i(\tau)}{\tau}\, .
\eea
Furthermore, using the first law of thermodynamics we get equation for the evolution of the entropy density:
\be\label{ds_i}
\frac{ds_i (\tau)}{d\tau} = -P_{esc} (\tau) s_i(\tau)
- \frac{s_i(\tau)}{\tau} +  \frac{\Phi_i(\tau)}{T_i(\tau) \tau}  \, .
\ee
Here the shear is defined to be positive, $\Phi = \pi^{00} - \pi^{zz}$.
In the standard first order theory, $\Phi \equiv \Phi_{1\rm{st}} = (4\eta)/(3\tau)$, where $\eta$ is the
shear viscosity coefficient.
In second order theory of relativistic fluid dynamics \cite{Israel_Stewart}, the initial value for the shear
and its evolution must be obtained independently.
Introducing the Reynolds number: $R = (e + p)/\Phi$, which represents the ration of non-dissipative to dissipative
quantities \cite{Baym:1985tn, Kouno}, then we can choose the initial value for
the stress such that $dR(\tau_0)/d\tau = 0$ \cite{Dumitru_2}.
Even in second order theory this physically motivated initial value for the stress brings
the behavior of the system very close to the standard first order theory during the evolution,
but in any case the $\Phi$ relaxes to the first order-theory, given by the relaxation time for the shear.
Therefore in the following we use the first order theory to describe the late stage of the
fluid dynamical evolution.

\section{Results and discussions} \label{Results}

\begin{figure}[hbt!]
\centering
\includegraphics[width=8cm, height = 5.5cm]{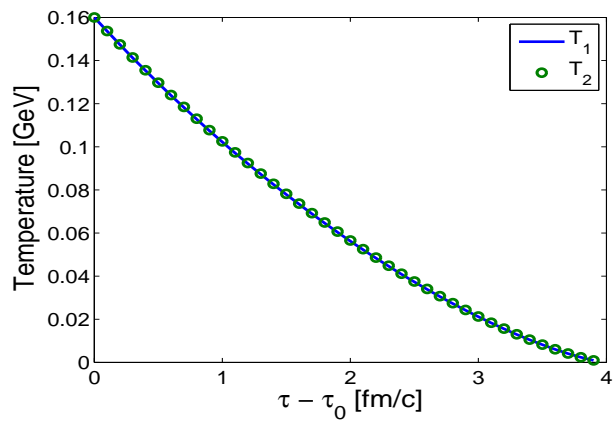}
\caption{The temperature of the interacting matter as a function of the freeze-out duration $\tau - \tau_0$.
$T_1$ is for a perfect fluid and $T_2$ is for a dissipative fluid.}
\label{fig_01}
\end{figure}

\begin{figure}[hbt!]
\centering
\includegraphics[width=8cm, height = 5.5cm]{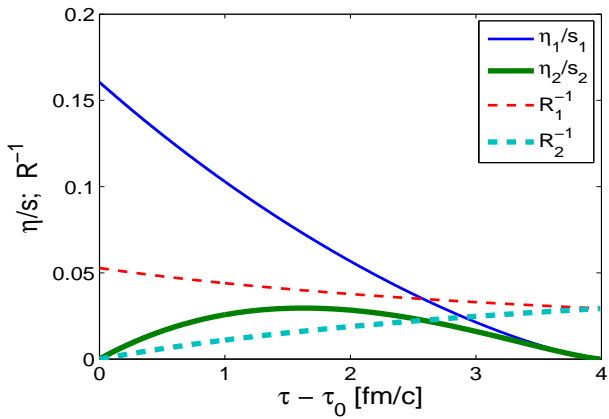}
\caption{The ratio of shear viscosity to entropy density,
and the inverse of the Reynolds number as a function of the freeze-out duration $\tau - \tau_0$;
see the text for more explanation.}
\label{fig_02}
\end{figure}

We assume that we deal with a massless pion fluid at freeze-out with an
EOS of an ideal fluid, $e(\tau) = 3 p(\tau)$, where the energy density is, $e = a T^{4}$,
$a = g_{\pi} \pi^2/30$, $g_{\pi}=3$ is the degeneracy of pions, while the entropy density is $s=4aT^3/3$.
The particle density is $n = bT^3$, where $b = g_{\pi} \zeta(3)/\pi^2$.
Here we recall the result of Gavin \cite{Gavin} for the shear viscosity coefficient, (see also \cite{HiranoGy} and references therein),
in case of a massless pion fluid,
calculated in the relaxation time approach from the relativistic Boltzmann transport equation,
\be\label{eta}
\eta(\tau) = \frac{4 \tau_{th}(\tau)}{15} e(\tau)\, .
\ee
%
The EOS together with eq. (\ref{de_i}) and eq. (\ref{eta}),
gives the following differential equation for the temperature,
\bea\label{temperature}
\frac{d T(\tau)}{T(\tau)} = - \frac{1}{3}P_{esc} (\tau) d\tau
- \left(\frac{1}{3} - \frac{ 4 \tau_{th}(\tau)}{45 \tau} \right) \frac{d\tau}{\tau} \, ,
\eea
leading to the following solution with the escape rate eq. (\ref{escape}) for fixed relaxation time, i.e.,
$\tau_{th}(\tau) = \tau_{th}(\tau_0)$,
\be\label{temp_1}
T(\tau) = T_0 \left[ \frac{\tau_0}{\tau}\right]^{\frac{1}{3}}
e^{\frac{4 \tau_{th}(\tau_0)}{45} \left(\frac{1}{\tau_0} - \frac{1}{\tau}\right)}
\! \left[\frac{L + \tau_0 -\tau }{L}\right]^{\frac{L}{3\tau'(\tau_0)}} \, ,
\ee
where $T_0 = T(\tau_0)$.
The result for a perfect fluid can be obtained in the limit when, $\tau_{th}(\tau)\rightarrow 0$.
Here we also take into account, with a simple ansatz, that during freeze-out the relaxation time increases as
the system dilutes and becomes sparser because of the expansion and the removal of interacting matter.
Thus even if the freeze-out starts from a perfect fluid, the relaxation time
increases during the finite time, $L$, such that:
$\tau_{th}(\tau) \rightarrow \tau_{th}(\tau_0)(\tau - \tau_0)/L$, reaching
the dissipative fluid limit, $\tau_{th}(\tau_0)$, at $\tau = \tau_0 + L$.
\\ \indent
For the purpose of illustration, and to obtain numerical results, we assumed the following initial values
for our parameters.
The continuous escape of particles lasts, $L = 4$ fm/c, measured from
the start of the freeze-out, $\tau_0 = 5$ fm/c.
We assume that the initial thermalization time and freeze-out characteristic time is
$\tau_{th}(\tau_0) \sim \tau'(\tau_0) \approx 1$ fm/c.
This is consistent with the average mean free time between collisions estimated from
two-particle correlations \cite{pion_mfp}.
The initial temperature is $T_0 = 160$ MeV.
\\ \indent
The evolution of the temperature is shown on Fig. \ref{fig_01}.
Here $T_1$ and $T_2$ show the decrease of the temperature of the interacting matter
during freeze-out from a perfect fluid and dissipative fluid respectively.
The latter corresponds to eq. (\ref{temp_1}), while $T_1$ corresponds to the
same equation in the perfect fluid limit.
The temperature decreases to zero in both cases signalling clearly the completed removal of
the interacting matter.
\\
\\
We have also employed the ansatz for an increasing relaxation time,
$\tau_{th}(\tau) = \tau_{th}(\tau_0)(\tau - \tau_0)/L$,
and checked our result against this approach; the temperature curve is essentially the same as earlier.
In all cases the viscosity is small and does not affect the fast decrease of the temperature
due to the freeze-out of matter, but in case of large viscosity the temperature decrease is slower
than in the perfect fluid limit.
\\ \indent
Fig. \ref{fig_02} shows the ratio of shear viscosity to entropy density and the inverse of the Reynolds number, using
a fixed and a dynamically increasing relaxation time.
From eq. (\ref{eta}) the ratio of shear viscosity to entropy density is:
$\eta(\tau)/s(\tau) = T(\tau) \tau_{th}(\tau)/5$.
Here $\eta_1/s_1$ shows the ratio of shear viscosity to entropy density for a fixed relaxation time.
Initially this value is close the conjectured minimal bound, $\eta/s \geq 1/(4\pi)$ \cite{Kovtun_1},
(see also \cite{HiranoGy, Csernai_1}, and in fluid dynamical simulations \cite{Romatschke, Romatschke2}).
In this case, $\eta_1/s_1$ is decreasing with the temperature.
The next, $\eta_2/s_2$, ratio employs the increasing relaxation time.
Since the mean free time between the collision increased so did the ability to carry momentum,
i.e., viscosity, and the ratio increases \cite{Csernai_1}.
Since we remove energy (momentum) from the system, the remaining cold matter has less energy (momentum) available
to transfer causing the decrease of the ratio.
\\
\\
The inverse of the Reynolds number is: $R^{-1}(\tau) =  4\tau_{th} (\tau)/(15 \tau)$.
Here, $R^{-1}_1$ shows the evolution of the ratio of dissipative to non-dissipative quantities with a
constant relaxation time.
$R^{-1}_2$ shows the transition from a perfect fluid to a dissipative fluid during freeze-out, where
the viscosity of the remaining interacting component is increasing with time.
\\ \indent
The next figure, Fig. \ref{fig_03}, shows the entropy production as a function of the freeze-out duration,
for a fixed relaxation time $S_1$ and for the increasing relaxation time $S_2$ respectively.
We can explicitly estimate the entropy production comparing the evolution of the entropy density, $s=4aT^3/3$,
of a viscous fluid to a perfect fluid and calculate the ratio of entropy densities, $s^{diss}/s^{perf}$, thus
for a fixed relaxation time,
\be
\frac{s^{diss}}{s^{perf}} =
\exp\left[ \frac{4 \tau_{th}(\tau_0)}{15} \left(\frac{1}{\tau_0} - \frac{1}{\tau}\right)\right] \, .
\ee
The entropy production is a function of the freeze-out duration and of the ratio of dissipative to
non-dissipative quantities, $R^{-1}$, but it is independent on the continuous removal of interacting matter.
Thus during the gradual transition from perfect to dissipative fluid,
taken into account through the increasing thermalization time, the total entropy produced at the end
is as much as during the evolution of a dissipative fluid undisturbed by the remove of interacting matter,
since the above ratio is the same with or without the freeze-out term.
\\
\begin{figure}[hbt!]
\centering
\includegraphics[width=8cm, height = 5.5cm]{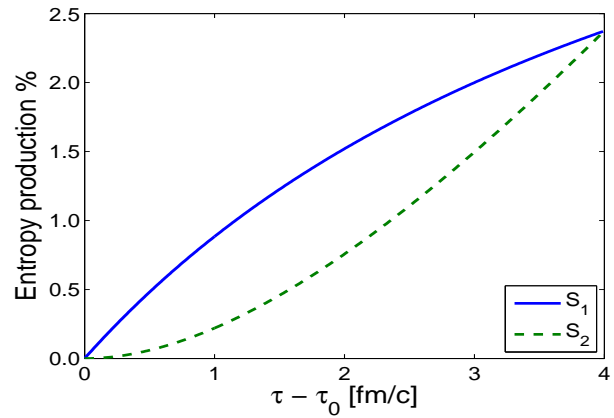}
\caption{The entropy production as a function of the freeze-out duration $\tau - \tau_0$.
Here $S_1$ and $S_2$ corresponds to the entropy production at the freeze-out from a dissipative fluid
with constant and increasing relaxation time respectively.}
\label{fig_03}
\end{figure}
%

\section{Conclusions}

We have incorporated a simple freeze-out criterion into the fluid dynamical model,
which continuously removes the conserved charge(s), energy, momentum and entropy from the system.
Since, entropy production is related to the speed of equilibration while the freeze-out drives
the system out of equilibrium, it is also important to include and treat the deviation from
equilibrium of the remaining interacting matter as this produces additional entropy which should
be added when estimating the total entropy produced in heavy-ion collisions \cite{Dumitru_2,Pratt,Shuryak}.
As long as these two processes are comparable and the deviation from equilibrium is small,
the entropy production is given by the standard viscous correction to the
equations of fluid dynamics.
\\ \indent
The result of our modest and illustrative analysis shows that the produced entropy is small even for long freeze-out times.
Compared to the entropy produced in a short time, i.e., $\simeq 1$ fm/c in the initial stage, during
the same amount of time at freeze-out this corresponds up to two orders of magnitude less entropy production.
Here one could use different relaxation times,  a different EOS and take into account higher order corrections
in the entropy four-current to (de-)increase the entropy production, but still in case of a massless pion
gas this would not be large.
Finally, taking into account the transverse expansion the entropy production due to
shear viscosity would slow down and by the time the longitudinal and transverse size of the
system become comparable, the shear viscosity vanishes \cite{Weinberg_book}.
However, even in this case substantial entropy production can be expected in a hadron gas where
the transport coefficients are different while the bulk viscosity and heat conductivity are
no longer negligible.
We believe that such calculation has much to clarify on the question of entropy production in high-energy
heavy-ion collisions.

%
\section*{ACKNOWLEDGMENTS}
The author would like to thank to A. Dumitru, D. H. Rischke and H. St\"ocker, for the useful comments
and discussions, and acknowledges support from the Alexander von Humboldt foundation.
%

\end{document}